\documentstyle[epsf,aps,twocolumn]{revtex}
\begin{document}
\draft
\title{Reconstructions of Ir (110) and (100): an ab initio study}
\author{Alessio Filippetti and Vincenzo Fiorentini}\address{
 Istituto Nazionale di Fisica della Materia and Dipartimento di 
Scienze Fisiche, Universit\`a di
Cagliari, I-09124 Cagliari, Italy}
\maketitle
\begin{abstract}
Prediction criteria for surface reconstructions are
discussed, with reference to ab initio calculations
 of the (110)-$1\times 2$ missing-row and (100)-$5\times 1$ 
quasi-hexagonal reconstructions
of Ir and Rh.
\end{abstract}
\pacs{PACS Nos: 0.0-x, 1.1+a}


The low index surfaces of most
 metals undergo symmetry-conserving multilayer relaxations
and, occasionally, surface reconstructions (either  symmetry-breaking
or -conserving), possibly accompanied by
a change in atomic density in the topmost layer.
Despite the good phenomenological characterization  of 
the reconstructed phases, a general picture of the driving 
mechanisms is still missing. To some extent,
it is not even clear if such a picture can be established at all.
Of course, different phases can always be compared in terms of the
respective free energies, but this implies a direct
calculation for the reconstructed phase.
A worthwile undertaking is therefore finding out which,
if any, computable quantities pertaining to the {\it unreconstructed}  
surface can be of help in rationalizing and possibly predicting 
instabilities of the ideal (relaxed) surface towards a reconstructed phase; 
in other words, whether or not surfaces can predict their own stability 
\cite{feib}. 
Obvious candidate indicators are surface energy, stress \cite{needs}, and 
strain \cite{dod}.
For instance,  the hex (100) reconstructions of Ir, Pt, and Au,
 whereby the top layer densifies into an hexagonal layer
laid on top of the (100) square lattice, has  been  modeled  in 
terms of the relief of surface stress \cite{fms}, whereas
faceting (i.e. a ``surface-energy driven'' transformation)
has been suggested as the origin of the $n\times$1 
reconstruction of  Au (110)
 \cite{binnig}. 

In this paper we take a first step towards the above goal by calculating the
 formation energies, relaxed geometries, and surface stresses 
of unreconstructed and reconstructed low-index surfaces of Ir
and Rh.
For both Ir  (100) and (110),  reconstruction is found
to be favorable. The same reconstructions for 
Rh are found to be  disfavored, in agreement with experiment;
results and technical details
for the ideal  Rh surfaces have been reported elsewhere
 \cite{ff}.
For Ir, we performed  local-density-functional-theory
\cite{dft} total energy and force calculations,
 using  iterative diagonalization in a parallel implementation 
\cite{valente},
smooth scalar-relativistic norm-conserving pseudopotentials,
   plane waves  cut off at 40 Ryd, k-points sets downfolded from
the 10-point fcc mesh, Fermi-surface smearing of
0.05 Ryd with the first-order Methfessel-Paxton
 approximation\cite{mp} for the occupation  function.
The resulting bulk parameters are $a_0=7.289$ bohr and $B = 4.2$
 Mbar.
Slabs are used encompassing 7 atomic layers
for the (100) and (111) surfaces, 9 layers for the ideal and
reconstructed (110), and 5 layers for the (100)-$5\times 1$.
The vacuum region  between slabs is equivalent to 5 atomic  layers 
(3 layers for the $5\times 1$).
All structures are relaxed according to
 Hellmann-Feynman forces  with a 
threshold of  1 mRyd/bohr $\sim$ 0.05 eV/\AA.\,
We found it convenient to use the method of Ref.\cite{chb} 
to accelerate the selfconsistency  convergence of forces,
and the method of Ref.\cite{fm} to calculate surface energies.

In Table \ref{t1} we list surface energies, relaxations, work
functions, and stress  of the three low-index (1$\times$1) surfaces of Ir.
They are in agreement with the usual picture for
cubic transition metal surfaces:  inward surface relaxations
occur,  increasing with surface roughness, while
workfunctions follow the opposite trend.
As pointed out previously \cite{fms}, many
   properties of Ir, and more generally the qualitative 
differences between 5$d$ and 4$d$ metals, can be traced back to 
 relativistic effects.
Despite the much larger core and in particular the larger $d$ shell,
 Ir (both atomic and crystalline) has about the same size as Rh.
This is due to the relativistic contraction of the  outer $s$ shell 
(caused by mass-velocity terms and $s-$core 
orthogonalization), having a 
pronounced maximum at nearby Au \cite{pykko}.
This  results in large bulk modulus, compact lattice, large
surface stress, and frequent surface reconstruction. 

We now come to describe the reconstructions.
Here we focus on Ir,  but qualitative conclusions 
also apply to Pt and Au. The missing-row reconstruction of
the (110) surface consists in the removal of one every two
rows of atoms along the [1$\overline{1}$0]  direction. 
The calculated reconstruction energy (the gain in energy upon 
reconstruction) 
is   $0.03$ eV/(1$\times$1 area), similar to the ab initio value
 of $0.05$ eV   for Au \cite{hb},
and to the semiempirical tight-binding result of $0.04$ 
eV for Pt \cite{toma}. The atomic relaxations 
(listed in Table \ref{t2}) are 
similar to those of the ideal (110) surface. The third-layer atoms 
 corresponding to the missing surface rows move upwards, those
neighboring  the remaining top rows  move downward, as expected.
The displacements tend to reduce the
surface  roughness, in accordance with the stress 
across the surface rows of the surface being tensile.
The components in the vertical direction are made to
vanish by relaxation. 

Does surface stress help in predicting this reconstruction ?
 We suggest that it does not: on the
 re\-con\-struc\-ted surface, the stress
(2.14 and 3.71 eV/atom in the two independent directions across and along
the surface troughs) is appreciably larger
than on the unreconstructed one
(1.70 and 3.21 eV/atom),  so stress relief cannot meaningfully
 be invoked in  this case. This is not too surprising at a closer look,
 since the reconstruction removes one top row, but exposes {\it
two additional} ((111)-like) rows in the process. Stress might be
removed from the topmost rows 
(although there is no way to tell this from just 
the integrated stress) but apparently it is just  transfered
to lower layers. This multilayer nature  of surface stress 
of fcc (110) is confirmed by preliminary results of a 
layer-by-layer decomposition of the surface stress,  to be presented in
detail elsewhere.

A more useful alternative is that of viewing the reconstructed 
surface as  a sequel of
adjacent (111) microfacets 
at alternate angles of about $\pm 35^{\circ}$ from the
vertical axis. While 
the upper and lower edge atoms will behave
differently from atoms on 
a clean (111) face,   it seems reasonable to argue that
the {\it average} local environment is that of a (111) surface.
This will be all the more true for a generic $n\times 1$
reconstruction in which $n-m$ every $n$ surface rows are removed
from the $m$-th layer ($n-1$ in the first layer, $n-2$ in the 
second,   down to 1 in the $n-1$-th layer), and all layers down
to the  $n+1$-th expose atoms at the surface.
It is then straightforward to predict 
 the reconstruction on the basis  of ideal-surface
quantities only. The  
surface energy per atom of the reconstructed surface 
is expressed as
\begin{equation}
\sigma^{rec}={\sigma_{(111)}\over A_{(111)}^{1\times1}}\>
{a_0\>\sqrt{a_0^2+d^2} \over \sqrt{2}}
\label{pippa}
\end{equation}
where $a_0$ is the bulk lattice costant, and $d$ the inter-layer distance.
For the unrelaxed structure $d= a_0/\sqrt{2}$ and
$\sigma_{rec}$ = 2 $\sigma_{(111)}$.  
With relaxation, $d$ gets  shorter and the surface energy
is lowered.
In the same spirit we can calculate the surface stress per cell 
by taking the stress per unit area equal to that of the the clean
(111) surface. The in-plane $[1\overline{1}1]$ component
(i.e. parallel to the chains) is simply obtained by
\begin{equation}
\tau^{rec}_{[1\overline{1}1]}=
{\tau^{surf}_{(111)}\over A_{(111)}^{1\times1}}\>
{a_0\>\sqrt{a_0^2+d^2} \over \sqrt{2}}
\label{pippa2}
\end{equation}
while the $[100]$ component in the relaxed configuration can be
calculated by projecting the planar component of $(111)$ surface
along the $[100]$ direction.

The faceting model gives a  reconstruction energy of 0.04 eV/atom 
respectively, which compares well with the ab initio value of 0.03;
the model also gives surface stresses
of  2.20  and 3.82 eV/atom
in the two independent surface 
directions, to be compared with ab initio values of  2.14 and 3.71  eV/atom.
 Two points are important: 
first,  using only (1$\times$1)-surface
quantities, the model predicts {\it directly} the
faceting of Ir (110) into (111) microfacets; second,
 it predicts accurately
 both surface energy and stress, at least in comparison with
ab initio values.
A further indication of the plausibility of the faceting picture
is that the work function {\it increases} by about 0.3 eV upon 
reconstruction: while this looks strange if one views the 1$\times$ 2
phase as just a rougher (110), this
 indicator of  surface smoothening 
fits naturally into the (111)-microfaceting picture.
Generally speaking, this picture is also in agreement with 
previous electronic kinetic-energy reduction  arguments \cite{hb}.
One more point to note is that the above arguments imply  that
the $n\times 1$ reconstruction should occur for any $n$. This
agrees with the simultaneous observation of $n\times$1 domains with
$n$ up to 4  for Au (110). On the other hand, the formation of 
 $n\times$1 domains implies the existence of domain walls
consisting of steps or step bunches of total heigth $n-1$, 
the cost of which  becomes rapidly too large for the surface
to  afford a large-$n$  reconstruction.

The (100)-$5\times 1$ reconstruction presents quite different features.
The atomic density  increases due to a contraction of the atomic rows
along the [100] direction, so that 6 atoms can be placed into 5 times a
$1\times 1$ cell area. The atoms are arranged in a buckled 
close-packed quasi-hexagonal structure. We calculate
surface energies, relaxations, and  stress for 
the ``two-bridge'' (TB)
and the ``top-center'' (TC) configurations \cite{bh}.
Both the TB and TC reconstructed 
configurations are found to be  favorable,
with a reconstruction energy of $0.14$ eV/(1$\times$1 area), 
and $0.11$ eV/(1$\times$1 area), respectively.
The TB configuration is the most favored, in agreement with
dynamical  LEED results \cite{bh}; 
the geometry of the hex layer agrees reasonably with
that predicted by LEED, although we obtain a smaller buckling.
The relaxation energies with respect to the experimental positions 
are about
 $0.1$ eV for both the configurations; also, Rh (100) is found not
to reconstruct (rec energy $\sim -0.1$ eV). These latter
 results   confirm further
 the conclusions of Ref.\cite{fms} on Pd and Pt, which were 
based on calculations for the experimental configuration.

Unlike the (110) missing-row, the ``quasi-hex''  reconstruction 
enhances the surface roughness,  as signaled by a decrease in the 
workfunction  of about 0.2 eV with respect to the (1$\times$1) phase,
in good agreement with the experimental drop
\cite{kup}  of $\sim$ 0.15 eV.
This behavior may seem unexpected for a transition to a
close-packed structure. However the quasi-hexagonal overlayer
matches poorly  the square (100) substrate, and 
is strongly buckled (by 15 to 40 \% of the 
ideal interlayer spacing). The
 net effect is a decrease of the surface  dipole.

The surface stress of the (1$\times$1) surface has been invoked as driving 
the $5\times 1$  reconstruction \cite{fms}. The lowering 
of in-plane symmetry causes an anisotropy of the surface stress;
the atomic density increases along the [110] direction, 
resulting in a large relief of surface stress ($\sim$ 40 \%).
 On the other hand, along [1$\overline{1}$0]  the 
first neighbor distance  remains
unaltered  on the reconstructed 
surface, and four new neighbors 
contribute to the stress 
along that direction. As a consequence, the stress along
[1$\overline{1}$0] is indeed found to increase 
(by about the same amount). 
A symmetry-breaking distortion, such as the domain rotations
occurring at high temperature, will allow further stress relief
 to set in along this direction.

In any event,
 a direct comparison between the in-plane stress of the
clean and  reconstructed surfaces is again not really helpful.
The integrated stress of the (1$\times$1) surface gives information about
the ``driving force'' (the in-plane stress  in the surface 
layer). This stress is relieved to  to some unknown extent
upon reconstruction.  The integrated stress in the reconstructed 
phase  contains additional contributions from bonds between the 
substrate and the mismatched overlayer which were not present in
the (1$\times$1) phase -- that is, it contains information about both
 the residual ``driving'' force, and the ``resisting'' forces, in
proportions that cannot be disentagled. Basically, as 
in previous studies, the problem
of extracting an independent value for the mismatch energy 
remains unsolved.
We suppose that a more useful reconstruction
indicator would be the surface strain of the (1$\times$1) phase
 as defined in Ref. \cite{dod};
this is in a way the strain effectively imposed onto
 the top layer by the infinite bulk
 to make the former fit onto the substrate. If the strain
 thus  determined is larger than a critical strain extracted from
e.g. elasticity theory, a transition behavior of the surface layer
 from a coherent (unreconstructed)
to an  incoherent  (reconstructed) state
can be  expected.
We are currently addressing this point and  will discuss it elsewhere.

It may be useful to note that there is a further basic
difference between the two reconstructions discussed above:
the density increases in the 5$\times$1 and decreases in the 1$\times$2.
The choice of the thermodynamic 
reservoir for atoms to be added ($5\times 1$)
or removed (1$\times 2$) from the surface is therefore essential. 
We used (quite naturally) the bulk chemical potential which
 is the most unfavorable choice for the $5\times 1$ reconstruction
in equilibrium with the bulk phase. With 
a ``less  expensive'' additional atom (for example, an isolated 
homoadatom), the reconstruction would be easier.
On the contrary, the same choice is most favorable
for  the (110) $1\times 2$ reconstruction. 

In conclusions, calculations for Ir and Rh surfaces support the view that no 
unique  quantity (energy, or stress, e.g.) can be invoked as a general 
driving force of
surface reconstruction for metals. We found evidence for microfaceting
of Ir (110). The results for reconstructed surfaces are 
in agreement with existing experiments.

\paragraph*{Acknowledgements -- } Stefano Baroni
and Riccardo Valente provided their parallel code, discussions
and technical assistance. Thanks to Dario Alf\`e for
code sharing. Most of the calculations were performed on 
the IBM SP2 of Centro Ricerche,
Sviluppo e Studi Superiori in Sardegna (CRS4), Cagliari, Italy,
within a collaborative agreement with Cagliari University.


\begin{table}[t1]
\centering\begin{tabular}{|l|rrr|r|r|rr|}
\hline
 &$\Delta d_{12}$  &$\Delta d_{23}$& $\Delta d_{34}$ &$\sigma$ &$W$ 
 &$\tau_{xx}^{\rm surf}$&  $\tau_{yy}^{\rm surf}$  \\
\hline\hline
(111)  &--1.3 &--0.2 & 0.0 & 1.31 & 5.92 &\multicolumn{2} {c}{1.96}\vline \\
\hline
(100)  &--3.8  & 1.0  &--0.5 &1.85 & 5.92 & \multicolumn{2} {c}{1.86}\vline \\
\hline
(110)  &--11.6  & 5.4 & --1.3  & 2.59 & 5.45 & 1.70 & 3.21\\
\hline
\end{tabular}

\caption[T1]{\footnotesize Inter-layer relaxations (percentage
variation with respect to ideal layer spacing), surface energy
$\sigma$ (eV), work function $W$ (eV) and surface stress $\tau$ (eV/atom)
for the three low-index clean surfaces of Ir. All values refer to
the fully relaxed structures.}
\label{t1}
\end{table}

\begin{table}[ht]
\centering\begin{tabular}{|l|rrrrr|rr|}
\hline
 & $\Delta z_1$ & $\Delta z_2$ & $\Delta z_{3a}$ & $\Delta z_{3b}$
 & $\Delta z_4$ & $\Delta x_2$ & $\Delta x_4$ \\
\hline
 Rh & --10.8 & --2.5 & --1.6 & 3.0 & --1.3 &--0.3 &--0.2 \\
\hline
 Ir & --10.3 & --1.9 & --2.5 & 5.2 & --0.8 &1.0 &--0.8 \\
\hline
\end{tabular}
\caption{ Relaxations of $1\times 2$ (110) reconstructed surface.
$\Delta z_n$ and $\Delta x_n$ are vertical and planar changes,
respectively, of the n-th layer atoms from ideal positions,
(\% of the ideal inter-layer spacing). For the 
third layer, two kinds
of shift occur (see text).}
\label{t2}
\end{table}



\begin{thebibliography}{99}

\bibitem{feib}
P. J. Feibelman, Phys. Rev. B {\bf 51}, 17867 (1995).

\bibitem{needs}
See e.g.  R. J. Needs, M. Mansfield, and M. J. Godfrey,
Surf. Sci {\bf 242}, 215 (1991).

\bibitem{dod}
B. W. Dodson,  Phys. Rev. Lett. {\bf 60},
2288 (1988).

\bibitem{fms}
V. Fiorentini, M. Methfessel, and M. Scheffler, Phys. Rev. Lett. {\bf 71},
1051 (1993).

\bibitem{binnig}
G. Binnig {\it et al.}, Surf. Sci {\bf 131}, L379 (1983);


\bibitem{ff}
 A. Filippetti, V. Fiorentini, K. Stokbro, R. Valente and
S. Baroni, in
{\it Materials Theory, Simulations, and Parallel Algorithms},
edito da E. Kaxiras and J. D. Johannopoulos,
MRS Proceedings {\bf 408}, 457  (1996).

\bibitem{dft}
 R. Dreizler and E. K. U. Gross {\it Density functional theory}, (Springer,
Berlin, 1990). The exchange-correlation energy  by
D. M. Ceperley and B. J. Alder, Phys. Rev. Lett. {\bf 45}, 566 (1980)
is used in the  parametrization of J. P. Perdew and A.  Zunger,
Phys. Rev. B {\bf 23}, 5048 (1981).

\bibitem{valente}
R. Valente and S. Baroni, to be published.

\bibitem{mp}
 M. Methfessel and A. P. Paxton, Phys. Rev. B {\bf 40}, 3616 (1989);
see also S. de Gironcoli, Phys. Rev. B {\bf 51}, 6773 (1995).


\bibitem{chb} 
C. T. Chan, K. P. Bohnen, K. M. Ho, Phys. Rev. B, {\bf 47},
4771 (1993).

\bibitem{fm}
V. Fiorentini and  M. Methfessel, J. Phys. 
Cond. Matt. {\bf 8}, 6525 (1996).

\bibitem{pykko}
See e.g.  P. Pykk\"o, Chem. Rev. {\bf 88}, 563 (1988). 

\bibitem{hb}
K. M. Ho and K. P. Bohen, Phys. Rev. Lett. {\bf 59}, 1883 (1987)

\bibitem{toma}
D. Tomanek, Phys. Lett. {\bf A 113}, 445 (1986).

\bibitem{bh} N. Bickel and K. Heinz, Surf. Sci. {\bf 163}, 
435 (1985)


\bibitem{kup}
J. Kuppers and H. Michel, Appl. Surf. Sci. {\bf 3}, 179 (1979)

\end{thebibliography}
\end{document}